\documentclass{jaa}

\usepackage[pdftex]{}
\usepackage{graphicx}

\begin{document}\sloppy


\title{Deep $V$ and $I$ CCD photometry of young star cluster NGC 1893 with the 3.6m DOT}
\author{Neelam Panwar\textsuperscript{1*}, Amit Kumar\textsuperscript{1,2}, and S. B. Pandey\textsuperscript{1}}
\affilOne{\textsuperscript{1}Aryabhatta Research Institute of Observational sciencES, Manora Peak, Nainital-263001, India\\}
\affilTwo{\textsuperscript{2}School of Studies in Physics and Astrophysics, Pandit Ravishankar Shukla University, Chattisgarh 492010, India}


\twocolumn[{

\maketitle

\corres{neelam@aries.res.in}

\msinfo{2021}{2021}


\begin{abstract}
        Young star clusters consisting of massive stars are the ideal sites to study the star formation processes 
	and influence of massive stars on the subsequent star formation. NGC 1893 is a young star cluster associated 
	with the H{\sc ii} region Sh2-236. It contains about five `O’-type stars and several early `B'-type stars. 
	It is located at a moderate distance of $\sim$ 3.25 kpc and has a reddening, 
	E(B-V) $\sim$ 0.4 mag. To characterize the young low-mass stellar population in the central portion of the cluster, 
	we carried out deep $VI$ band observations of the region using the 4K$\times$4K CCD $IMAGER$ mounted on the 3.6-m Devasthal Optical Telescope. 
	Our analysis shows that the present data are deep enough to detect stars below V$\sim$ 24 mag. 	We found optical counterparts 
	of $\sim$ 220 candidate members, including young stars and unclassified cluster members from Caramazza et al. (2008). We estimated the membership probabilities of the Gaia sources (mostly bright stars with G$<$ 19 mag) 
	located within the cluster radius using the Gaia EDR3. 
	Toward the fainter end, we used the optical color-magnitude diagram (CMD) to select the cluster members from a sample of young stars. 
	The locations of young stars on the CMD show that a majority of them are low-mass stars with age $<$ 10 Myr. 
	The unclassified member candidates and X-ray sources from Caramazza et al. (2012) are also found to be young low-mass stars. 
	In total, we identified $\sim$ 425 young stars with age $<$ 10 Myr, and 110 of these are new. Most of these stars appear kinematic members of the cluster. 
	By examining the CMD for the stars in the cluster region, we suggest that the cluster has insignificant contamination due to field stars in the pre-main sequence zone of the CMD. The slope of the mass function in 
	the mass range 0.2$\le$M/M$_\odot$$\le$2.5 is found to be $\Gamma$ = -1.43 $\pm$ 0.15, consistent with those of other star-forming complexes. The spatial distribution of the young stars as a function of mass suggests that toward the cluster center, most of the stars are massive. 
	\end{abstract}
}]


\doinum{}
\artcitid{\#\#\#\#}
\volnum{000}
\year{2021}
\pgrange{1--}
\setcounter{page}{1}
\lp{1}

\section{Introduction}
Stars form in the molecular clouds with a variety of masses, ranging from high-mass ($>$ 8 M$_\odot$) 
to the low-mass upto 0.08 M$_\odot$ and most of the stars emerge as clusters (Lada \& Lada 2003). Despite several observational and theoretical works focusing on the processes involved in the formation and evolution of stars and star clusters, 
a number of issues persist. As young star clusters possess young stars of a diverse mass range,  formed from the same molecular cloud, they 
are particularly suited to enhance our understanding on the physical processes related to the star formation, such as, whether star 
formation is a fast or slow process, what is the shape of the initial mass function (IMF) towards the low-mass end, and the 
total star formation efficiencies. 
However, the presence of massive stars in such systems may significantly influence 
the evolution of low-mass stars and subsequent star formation. As soon as the massive stars form, 
they tend to ionize the natal cloud and create expanding H{\sc ii} region. The expanding edge of the 
H{\sc ii} region (I-front) interacts with the surrounding cloud and may trigger star formation via 
various processes. Thus, young star clusters associated with the H{\sc ii} regions and young stellar objects (YSOs) are ideal sites to study the 
influence of massive stars on the formation and evolution of low-mass stars and the processes involved 
in triggering star formation. 

Observations of the early stages of star formation, i.e., Class~{\sc 0}/{\sc i}/{\sc ii} YSOs show that these are associated with the disk accretion, jets and outflows. 
However, jets and outflows are only present when a young star possesses an accretion disk. The strength of a jet depends on the evolutionary status of the driving source and the less evolved source 
(with thick circumstellar disk) has more powerful outflows. Hence, the presence of jets and outflows in a star forming region can be used to identify young stars 
that are deeply embedded in the molecular clouds. 

As YSOs possess circumstellar accretion disks during their earlier stages of formation, 
they also exhibit excess emission in the longer (particularly IR) wavelengths. However, the amount of excess emission depends on different evolutionary classes, i.e., Class~{\sc 0}, Class {\sc i}, Class {\sc ii}, and 
Class {\sc iii} (Andr{\'e} 1995). Hence, in general, this property of the young stars is used to identify and classify them 
(Lada et al. 2006; Luhman 2012). 
As Class~{\sc iii} objects are disk anemic or possess thin disks, they present no or little excess emission in infrared (IR). Therefore, IR observations become less sensitive for the identification of Class {\sc iii} objects. However, X-ray observations can be utilized to obtain a complete census of young stars 
as both Class {\sc ii} and Class {\sc iii} objects are generally 
more luminous in the X-ray wavelengths compared than their main-sequence (MS) counterparts (e.g., Feigelson \& Montmerle 1999; Getman et al. 2012).
Hence, IR and X-ray observations are widely used to identify and characterize a complete census of YSOs in star-forming regions.

{\hspace{-0.65cm}\bf Our target: NGC~1893}\\
NGC~1893 is a young star cluster located in the Auriga OB2 association. 
It is associated with the H{\sc ii} region W8 (or Sh2-236) and contains about five `O’-type stars and 
several `B'-type stars (Marco \& Negueruela 2002). Recent studies place the cluster at a moderate distance ranging from $\sim$ 3.2 - 3.6 kpc. 
Reddening, E($B$ - $V$), toward the cluster direction ranges from 0.4 to 0.6 mag (Sharma et al. 2007, Prisinzano et al. 2011, Lim et al. 2014). 
The radial extent of the cluster is found to be $\sim$ 6$^\prime$ (Sharma et al. 2007). 
Using different color excess ratios over a wide wavelength range, Lim et al. (2014) found a normal reddening law (R$_V$ = 3.1) towards the cluster.
CO (1-0) observations of the H{\sc ii} region/ cluster complex show that it still consists of molecular gas having V$_{LSR}$ = ($-$7.2 $\pm$ 0.5) km $s^{-1}$ (Blitz et al.  1982). 
As NGC~1893 is located in the outer Galaxy and probably formed 
in the low-metallicity environment, it is interesting to probe how such a rich star cluster formed there despite 
of the expected unfavourable conditions for star formation in the outer Galaxy.

The region is associated with two tadpole nebulae, Sim 129 and Sim 130, and several YSOs (Maheswar et al. 2007, Caramazza et al. 2008, 2012, Prisinzano et al. 2011, Lim et al. 2014). 
The spatial distribution of the YSOs shows that they have an elongated and aligned 
distribution from the cluster center to Sim 129 and Sim 130. An age gradient from the massive stars of the cluster 
toward the nebulae is also observed (Pandey et al. 2013, Kuhn, Getman \& Feigelson 2015).  
Lata et al. (2012) identified 53 young stellar variables based on the $VI$ bands time-series photometric observations. 
They found 
that the rotational period of young stars decreases with stellar age and mass, and the amplitude 
in the light curves also declines with the same physical quantities. Based on a cumulative age distribution of 
the classical T-Tauri stars (CTTSs) and weak-line T-Tauri stars (WTTSs), Pandey et al. (2013) found 
that these are coeval. Hence, the cluster/ H{\sc ii} region complex is an active site of star formation and an ideal target to study the 
formation and evolution of the young stellar population. 

Though the cluster and associated H{\sc ii} region is a target of various optical photometric studies, most of these were shallow 
(V$\sim$ 21-22 mag) and complete upto $\sim$ 1 M$_\odot$ (Lim et al. 2014). As low-mass stars out-number high-mass stars in a star cluster, 
information on the low-mass stellar population of the young clusters is essential to infer the cluster properties, star formation histories  
and mass function (MF). Deep photometric observations are important tools to probe the faint low-mass stars and study the properties of young clusters.  

To characterize the low-mass stars and study star formation in the cluster region, we carried out deep $VI$ band 
observations of the region with the 4K$\times$4K CCD $IMAGER$ mounted on the 3.6-m Devasthal Optical Telescope (DOT). Our analysis show that 
the present optical data are $\sim$ 3 mag deeper than that of the previous studies (e.g., Sharma et al. 2007, Lim et al. 2014). 
\begin{figure*}
\centering
	\includegraphics[scale = 0.68, trim = 0 0 0 10, clip]{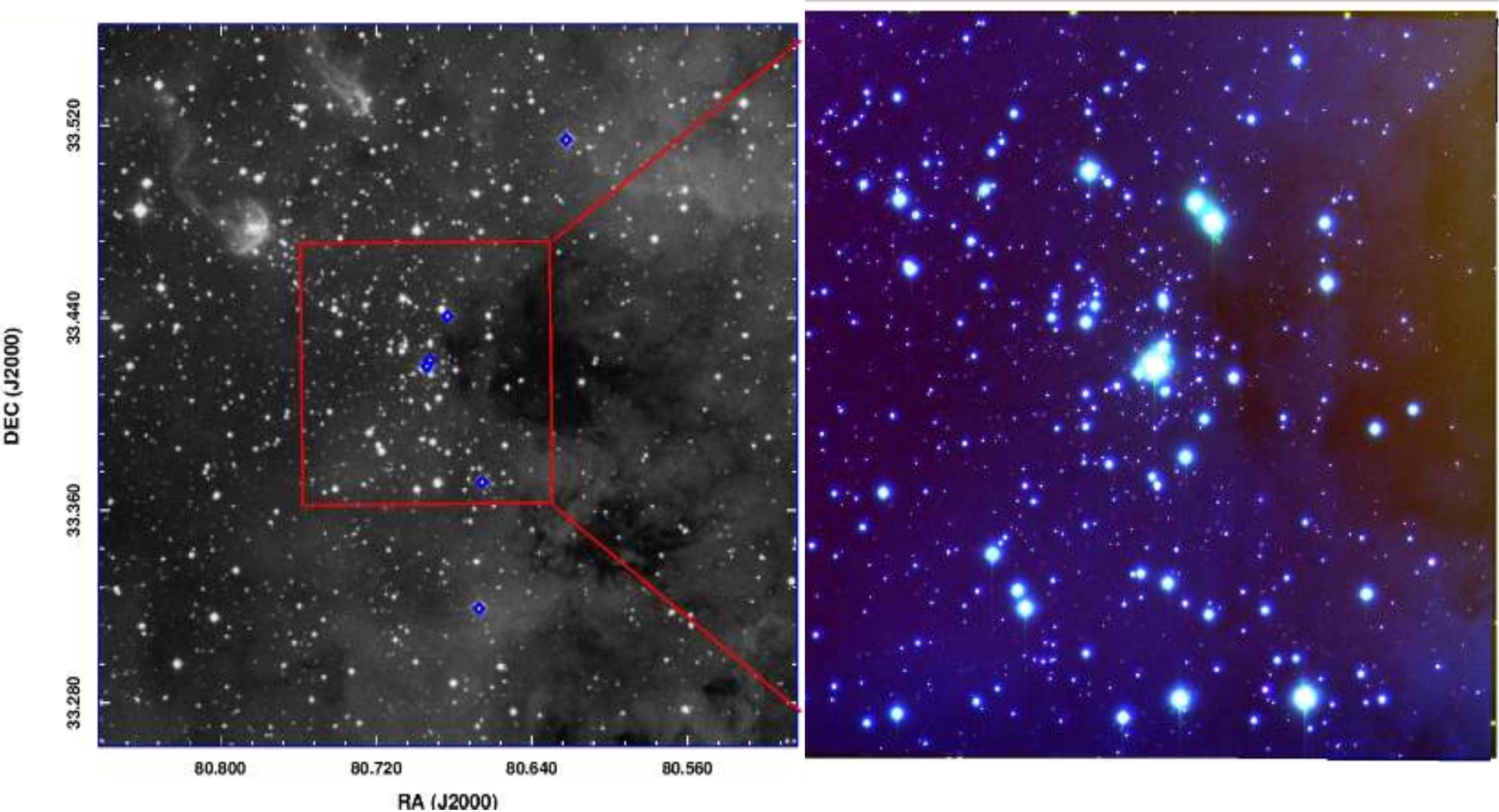}
	\caption{Left panel: DSS2-R band image of the NGC 1893 complex. Diamond symbols represent the locations of `O' type stars 
	in the region. Right panel: A color-composite view of the central portion ($\sim$ 6$^\prime$.5 $\times$ 6$^\prime$.5 FOV) of the cluster constructed using the DSS2-R (blue), 4K$\times$4K CCD IMAGER V-band (green) and I-band (red) images.}
        \label{fig1}
\end{figure*}
\section{Observations \& Data Reduction}
The $V$$I$ bands photometric observations of the central portion of the cluster NGC~1893 were carried out on 2020 October 15 using the 
4K$\times$4K CCD $IMAGER$ mounted at the axial ports of the 3.6m DOT (Pandey et al. 2018). 
With a plate scale of $\sim$ 0.095 arcsec/ pixel, the CCD covers a field of view of 6$^\prime$.5 $\times$ 6$^\prime$.5.
The observations were carried out in a 2$\times$2 binning mode to enhance the signal to noise ratio. 
The images were taken in readout noise and gain mode of 10 e$^{-}$/s and  5 e$^{-}$/ADU, respectively. 
The sky conditions were excellent throughout the night, and during the observations the average full-width half-maxima (FWHM) of the point sources was $\sim$ 0$^{\prime\prime}$.5. 
To avoid the saturation and contamination to the flux of the stars by the nearby bright stars, we took 12 exposures of 
300s and 210s in both $V$ and $I$-band, respectively. Thus, the total integration time for the $V$ and $I$ band images were 1 hour and 42 minutes, 
respectively. Along with the object frames, several bias and flat frames were also taken during the same night. 
Pre-processing of the object frames (i.e., bias subtraction, flat-fielding, etc.) was done using the $IRAF$ 
data reduction package. We combined the $V$ and $I$ band images separately using the $imcombine$ task of $IRAF$. 
We used the $DAOPHOT - II$ software package (Stetson 1987) for the source detection and the photometric measurements of the detected sources 
in the combined images. The point-spread function (PSF) was obtained for each frame using several 
uncontaminated stars and the $ALLSTAR$ task is used to obtain the instrumental magnitudes of the stars in each frame. 

The instrumental magnitudes were calibrated using the photometric measurements of the stars in the cluster NGC 1893 by Sharma et al. (2007). For calibration, we restricted the Sharma et al. (2007) catalog to 18$<$V mag$<$20 to avoid the effects of large 
photometric errors and saturation of bright stars towards the fainter and brighter ends, respectively. In total, we detected about 1750 stars in both $V$ and $I$ bands. Out of these, about 1510 stars have magnitude uncertainty $<$ 0.1 mag in both $V$ and $I$ bands. 

Prisinzano et al. (2011) also carried out optical and infrared (IR) observations of the young cluster NGC~1893 using Device Optimized for LOw RESolution, 
a 2K $\times$ 2K CCD camera with a plate scale of 0$^{\prime\prime}$.25 /pixel, mounted on the 3.6-m Telescopio Nazionale Galileo (TNG) at La Palma. 
They have reported seeing of $\sim$ 1$^{\prime\prime}$ during the observations.  
Though our observations are performed with a similar size telescope, the excellent atmospheric conditions (FWHM $\sim$ 0$^{\prime\prime}$.5) and better spatial 
resolution of 4K$\times$4K CCD $IMAGER$ are the advantages. We detected sources below $V$ $\sim$ 25 mag, however considering magnitude uncertainty of $<$ 0.1 mag in both 
$V$ and $I$ band, our photometry is limited to $V$ $\sim$ 24 mag. These values are comparable to $VI$ photometry of the same region 
taken  with  the TNG (see Figure 4 and Figure 8 of Prisinzano et al. 2011). Here, we note that during the observations, the reflectivity of the primary mirror of the 3.6m DOT was $\sim$ 60\%. Hence with 90\% reflectivity, we expect to reach about 0.5 mag deeper than the present observations.

Fig. 1 (left panel) shows the DSS2-R band image of the cluster NGC~1893 along with the Sim 129 and Sim 130 (toward the north-east). 
Diamond symbols represent the locations of massive `O'-type stars taken from SIMBAD. The square represents the area covered with the 
4K$\times$4K CCD $IMAGER$ observations (shown in right panel). A color-composite image of the central portion of the cluster (FOV $\sim$ 6$^\prime$.5 $\times$ 6$^\prime$.5) constructed using the DSS2-R (blue), 4K$\times$4K CCD $IMAGER$ $V$(green) and $I$-band (red) images is shown in the right panel of Fig. 1. 
\section{Results \& Discussion}
\subsection{Young stellar population in the region}
The spatial distribution of the young stars traces the sites of recent/ ongoing star formation. Hence, the identification and characterization of young stars in a region is 
essential to infer the star formation histories and the physical properties of the star clusters and H{\sc ii} regions. 
Caramazza et al. (2008) identified 359 YSOs in the cluster NGC 1893 using {\it Spitzer}-IRAC observations. Based on the IRAC color-color 
diagram and X-ray emission, these YSOs were characterized as Class~0/{\sc i}, Class~{\sc ii} or Class~{\sc iii} sources. Out of 359 YSOs, 
seven sources were classified as Class~0/{\sc i}, 242 Class~{\sc ii} and 110 Class~{\sc iii}. Caramazza et al. (2008) 
also reported $\sim$ 460 candidate members that were not detected at 5.8 and 8.0 $\mu$m 
and have 
[3.6] - [4.5] $>$ 0.3 mag. Therefore, they could not categorize these candidate members as YSOs. 
Prisinzano et al. (2011) identified 1034 disk bearing (Class {\sc 0}/{\sc i}/{\sc ii}) sources and about 442 diskless (Class~{\sc iii}) 
sources in the young cluster NGC 1893 by applying the Q-index method on the data of Caramazza et al. (2008). However, some of their YSO candidates 
($\sim$ 170 sources) appeared older than the majority of young population, which they attributed due to the edge-on disks and accretion activity. Caramazza et al. (2012) 
studied the coronal properties of the young stars in the NGC~1893 region using the $Chandra$ X-ray observations.

\subsection{Optical counterparts of the YSOs candidates}
As young stars possess disks in their early stages and show excess emission in the longer wavelengths, the estimation of the physical properties 
of the young stars using longer wavelength observations can be problematic. Since, optical observation mostly traces the 
photospheric emission from the young stars, these are the best suited to study the physical properties, such as age, mass, etc. of the young stars. 
In the 
present work, we have used the catalog of Caramazza et al. (2008) to guide our analysis and, in particular, to identify the young stellar 
population in the central portion of the cluster. 
We looked for the optical counterparts of the candidate members (Table 4 of Caramazza et al. 2008) in our optical data using a matching radius of 0.6 arcsec. We found optical counterparts of 226 member stars within the central portion ($\sim$ 6$^\prime$.5 $\times$ 6$^\prime$.5) of the cluster. 
Out of these, one source is Class~{\sc 0}/{\sc i} candidate, whereas 106 sources are Class~{\sc ii} and 42 sources are Class~{\sc iii} in nature. 
Though the rest of the 77 sources were designated as candidate members by Caramazza et al. (2008), these were detected in only 3.6 $\mu$m 
and 4.5 $\mu$m $IRAC$ bands 
and hence were not identified as YSOs based on the IRAC color-color diagram. Accordingly, the nature of these 
sources remained ambiguous. 
We looked for these 77 candidates in the YSO catalog of Prisinzano et al. (2011) and found that 55 of these 
were not present in their YSO catalog either. Since, we are using deep optical $VI$ photometry of the cluster in the present work, it will be helpful to examine the nature of these 55 candidate members of Caramazza et al. (2008). 

We also used the catalog of X-ray sources of Caramazza et al. (2012) to select the additional YSO candidates in the cluster. Using a matching radius of 
0.6 arcsec, we found optical counterparts of 336 X-ray sources. Out of these, 109 sources were common to those in the catalog of Caramazza et al. (2008). In summary, we found that the evolutionary status of $\sim$ 123 X-ray sources and unclassified candidate members were not previously known.   

\subsection{Kinematic members of the cluster NGC~1893 with Gaia EDR3}
The Gaia mission has revolutionized the kinematic studies of the stars and star clusters by providing rich astrometric 
and photometric data. In the present work, we used Gaia Early Data Release 3(EDR3; Gaia Collaboration et al. 2020) 
catalog to elucidate the kinematics of the stars in the young star cluster NGC~1893. We use Gaia EDR3 data for the sources within the cluster region (radius $\le$ 6 arcmin). 
We used the proper motion in RA ($\mu$$_\alpha$$^\star$) and proper motion in declination ($\mu$$_\delta$) for the stars within the cluster radius 
to generate the vector point diagram (VPD), where $\mu$$_\alpha$$^\star$ $\equiv$ ${\mu}_{\alpha}\cos(\delta)$.
We restricted only those sources with proper motion uncertainty less than 0.5 mas yr$^{-1}$ and G-band magnitude uncertainty less than 0.1 mag. 
The VPD for those sources is shown in Fig. 2. The over-density of sources can be easily noticed in Fig. 2. The proper motion of the stars in the cluster region 
peaks at $\mu$$_\alpha$$^\star$,$\mu$$_\delta$ $\sim$ -0.30, -1.43 mas yr$^{-1}$.   
A circular area of radius 0.7 mas yr$^{-1}$ around the peak of the over-density in the VPD is used to select the probable cluster members, and 
the remaining sources in the VPD are considered as field stars. 

To determine the membership probability of the stars in the cluster region, we used the approach discussed in Pandey et al. (2020). 
Assuming a distance of $\sim$ 3.25 kpc (Sharma et al. 2007) and a radial velocity dispersion of 1 km s$^{-1}$ for open clusters (Girard et al. 1989), 
a dispersion ($\sigma$$_{c}$)  of $\sim$ 0.06 mas yr$^{-1}$ in the PMs of the cluster can be expected. We calculated $\mu$$_{xf}$ = 0.84 mas yr$^{-1}$, 
$\mu$$_{yf}$ = -2.70 mas yr$^{-1}$, $\sigma$$_{xf}$ = 2.29 mas yr$^{-1}$ and $\sigma$$_{yf}$ = 3.76 mas yr$^{-1}$ for the probable field members. These values are further used to construct 
the frequency distributions of the cluster stars (${\phi }_{c}^{\nu }$) and field stars (${\phi }_{f}^{\nu }$) by using the equation given in Yadav et al. (2013) 
and then the value of membership probability for the $i$$^{th}$ star is calculated using the equation given below:

\begin{equation}
P_\mu(i) = {{n_c\times\phi^\nu_c(i)}\over{n_c\times\phi^\nu_c(i)+n_f\times\phi^\nu_f(i)}}
\end{equation}
\begin{figure}
\centering
\includegraphics[scale = 0.67, trim = 0 0 0 0, clip]{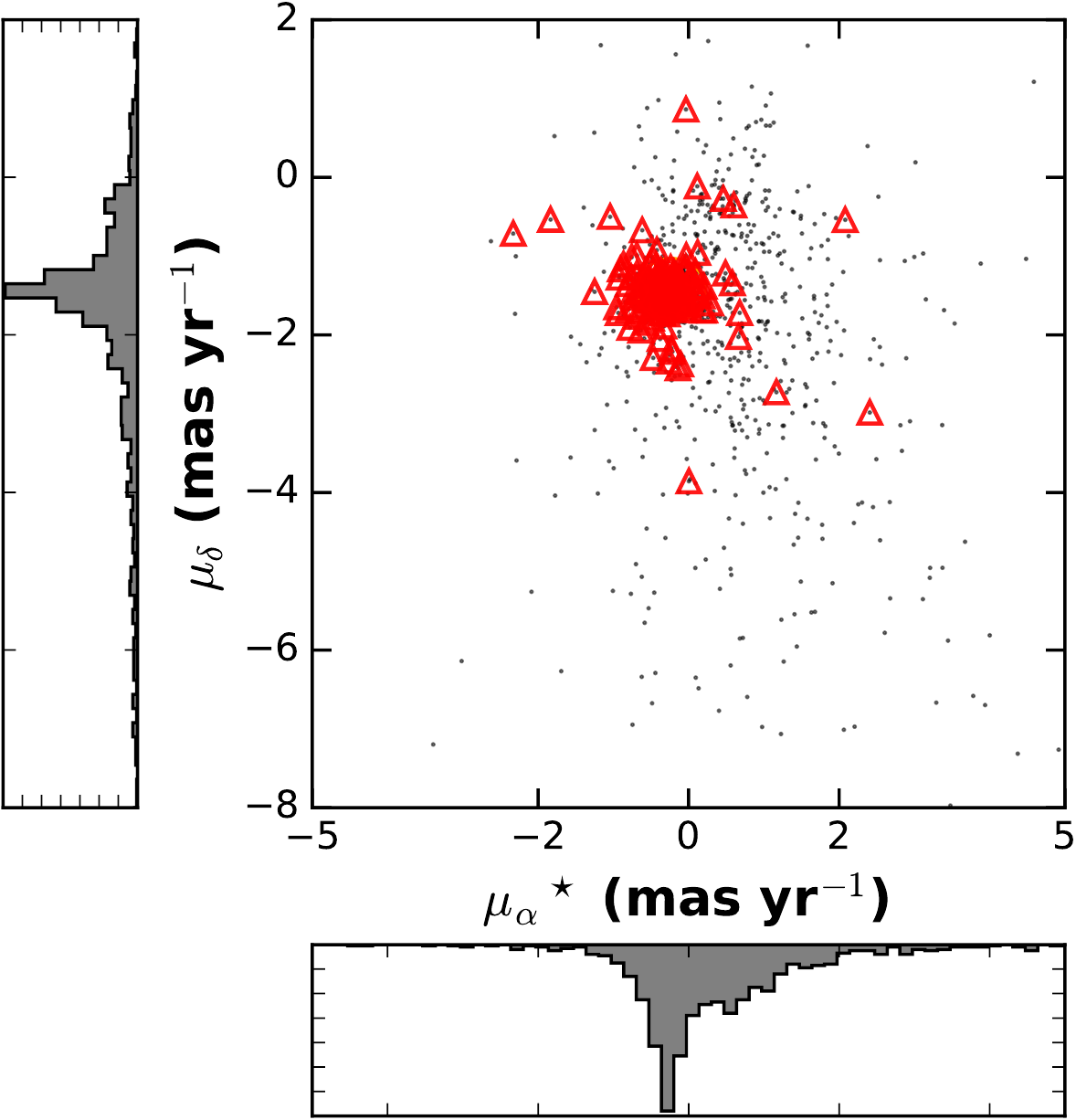}
	\caption{Proper motion vector-point diagram for the Gaia sources located within the cluster radius. The red triangles show the young stellar candidates having $VI$ 
	photometry in the present work.}
\label{fig2}
\end{figure}
\begin{figure}
\includegraphics[scale = 0.5, trim = 0 0 0 0, clip]{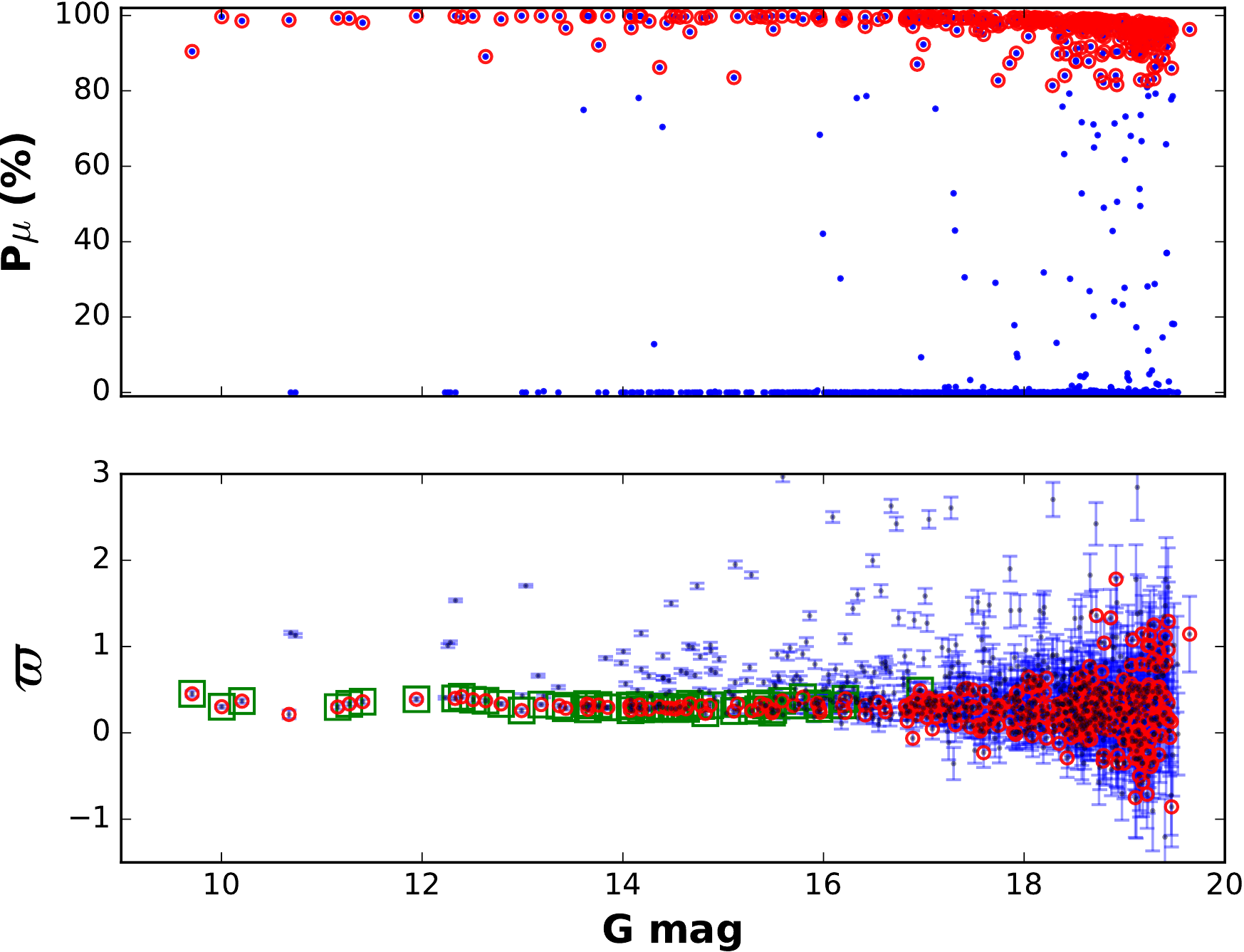}
	\caption{Top Panel: Membership probability for the stars located within the cluster area (radius $\sim$ 6$^{\prime}$) plotted as a function 
	of G-band magnitude. Bottom panel: The blue dots show the distribution of 
	parallax values for the Gaia sources within the cluster as a function of G magnitude. The error bars show respective uncertainties in the parallax values. 
	 The red circles represent the probable cluster members (P$_\mu$ $>$ 80\%). The green squares represent the probable cluster members with 
	 good parallax estimates ($\varpi$/$\sigma$$_\varpi$ $>$ 5).} 
\label{fig3}
\end{figure}

\begin{figure*}
\centering
\includegraphics[scale = 0.8, trim = 0 0 0 0, clip]{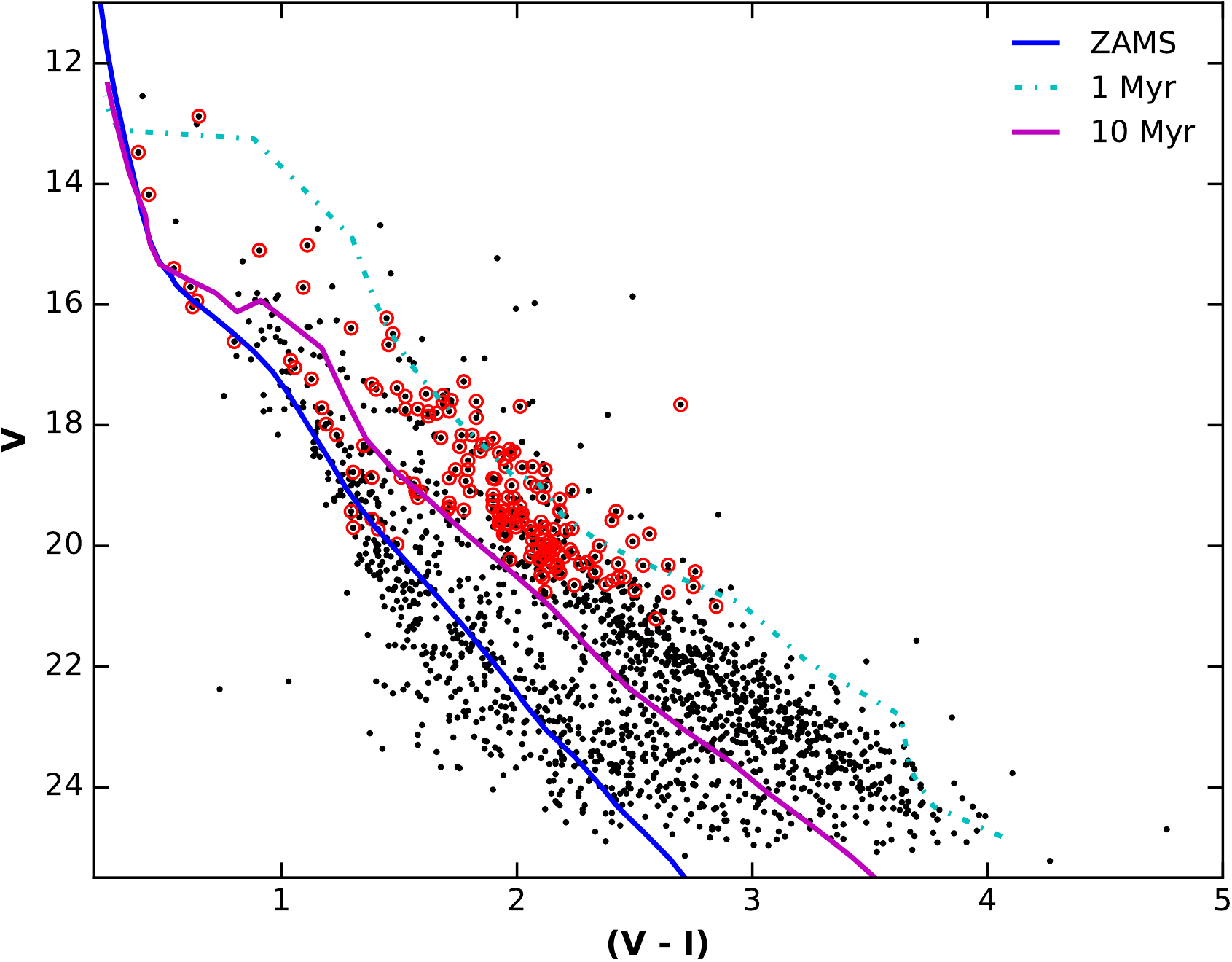}
\includegraphics[scale = 0.8, trim = 0 0 0 0, clip]{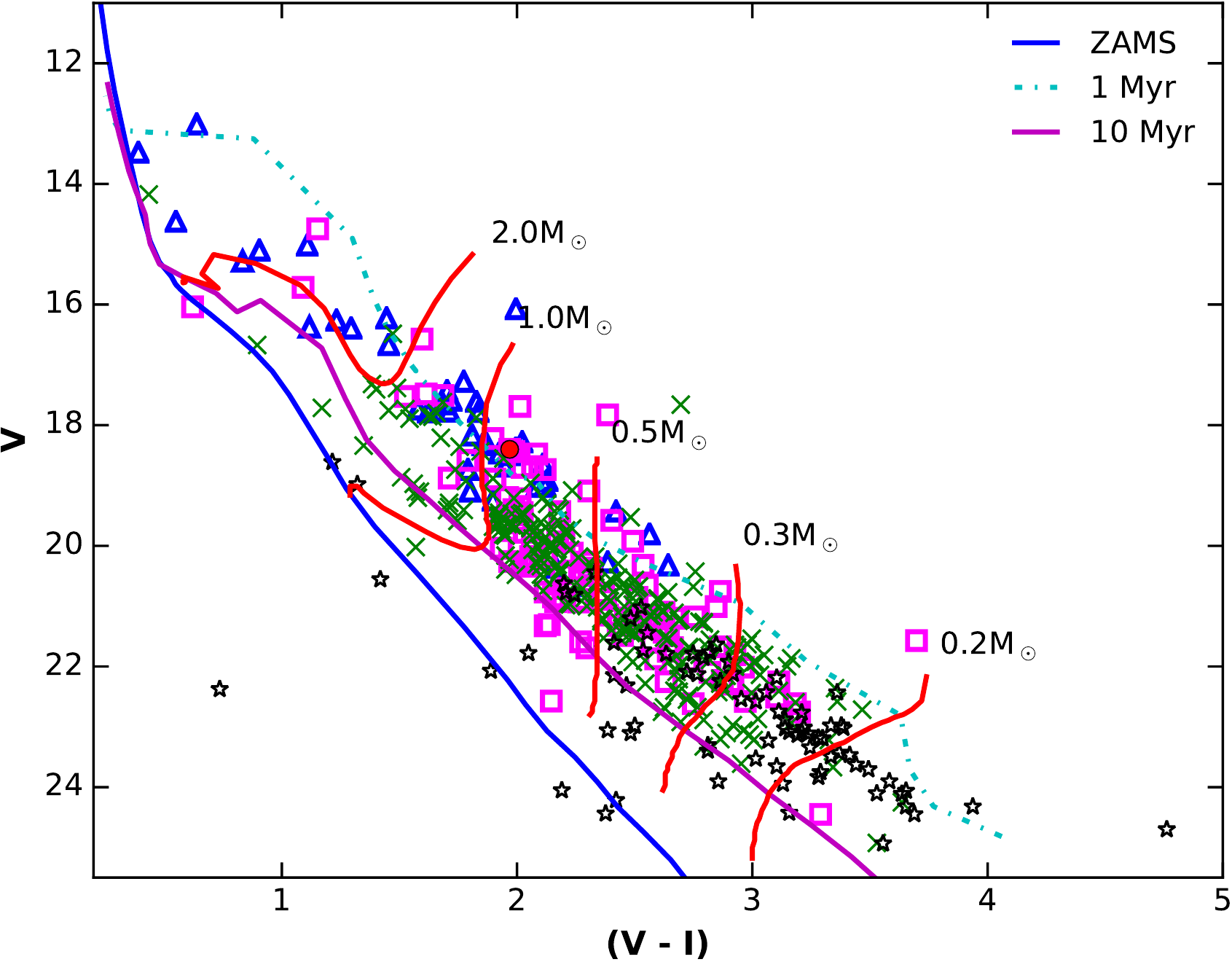}
	\caption{Top Panel: $V$/($V$-$I$) color-magnitude diagram for the stars in the central portion of the cluster ($\sim$ 6$^{\prime}$.5 $\times$ 6$^{\prime}$.5) obtained using the 4K$\times$4K CCD $IMAGER$ observations. The open circles represent Gaia sources with P$_\mu$$>$80\%.  
	Bottom panel: $V$/($V$ - $I$) color-magnitude diagram for the young stars in the cluster region. The red filled circle represents the 
	Class~{\sc i} object, magenta squares represent the Class {\sc ii} objects, and blue triangles 
represent the Class {\sc iii} objects from Caramazza et al.(2008). Cross symbols represent the optical counterparts of the X-ray sources detected by the $Chandra$ 
	observations (Caramazza et al. 2012) that were not listed in Caramazza et al.(2008). Star symbols represent 
	the candidate members of the cluster from Caramazza et al. (2008) that were detected in only two $IRAC$ bands and do not exhibit 
	X-ray emission. The zero-age main sequence (Girardi et al. 2002) and 1, 10 Myr pre-main sequence isochrones (Siess 
	et al. 2000) corrected for the adopted distance and reddening are also plotted.}
        \label{fig4}
\end{figure*}

where n$_c$ and n$_f$ are the normalized number of probable cluster members and field members, respectively. 
In figure 3 (upper panel), we have plotted the estimated membership probability for all the Gaia sources within the cluster radius as a function of G-band magnitude as blue dots. 
Gaia sources with high membership probability (P$_\mu$ $>$ 80\%) are shown with red circles. There seems to be a clear separation between the cluster members and field stars toward the brighter part, supporting the effectiveness of this technique. A high membership probability extends down to G $\sim$ 19 mag, whereas toward the fainter limits, the probability gradually decreases. 
A majority of the stars with high membership probability follow a tight distribution in the VPD. From the above analysis, we calculated the membership probability of 
$\sim$ 950 stars in the cluster region. 

We also plotted the parallax for all the Gaia sources as a function of G-band magnitude (dots in Fig. 3, bottom panel). The respective uncertainties in the parallax values are shown with the error bars. The red circles represent sources having membership probability 
P$_\mu$$>$ 80\%. We also estimated the cluster distance using the parallax values of the cluster members having high membership probability (P$_\mu$$>$ 80\%) and 
good parallax accuracies ($\varpi$/$\sigma$$_\varpi$ $>$5). These sources are shown with the square symbols in Fig. 3 (bottom panel). The median parallax value of these sources 
is 0.318 $\pm$ 0.054 mas. We estimate the cluster distance after correcting the median parallax value for the known parallax offset of $\sim$ -0.015 (Statssun \& Torres 2021). 
The distance estimate for the cluster using Gaia data comes out to be $\sim$ 3.30 $\pm$ 0.54 kpc, which is in agreement with that reported by Sharma et al. (2007).  

We also cross-matched the  YSOs that have counterparts in our optical photometry to sources in the Gaia catalog. Using a matching radius of $\sim$ 1.2 arcsec, we found $\sim$ 135 Gaia counterparts. These sources are shown with the red triangles in the VPD. Except for few contaminants, most of them follow the distribution of probable cluster members in the VPD, suggesting that these YSO candidates 
are the cluster members. 

From the above analysis, it is clear that using the Gaia EDR3 data, our sample of kinematic members of the cluster is restricted to only relatively bright members (G$<$ 19 mag). Therefore, to select the candidate members towards the low-mass 
end, we adopted the procedure discussed in the following section. 
\subsection{Color-Magnitude Diagram: Constraining Young Nature of the Candidate Members}
The color-magnitude diagram (CMD) is often used to study the mass distribution and evolutionary stages of the stars of a cluster. 
In Fig. 4 (top panel), we show the $V$ / ($V$ - $I$) CMD for all the stars in our optical catalog within the cluster region. 
The red circles represent the cluster members with high membership probabilities (P$_\mu$ $>$ 80\%). We have also plotted a zero-age main sequence (ZAMS) isochrone (thin blue curve) from Girardi et al. (2002) and 1, 10 Myr pre-main sequence (PMS) isochrones (dashed and thick curves, respectively) from Siess et al. (2000). In the present work, we adopted a distance of $\sim$ 3.25 kpc 
and reddening E($B$ - $V$) of 0.4 mag for the cluster (Sharma et al. 2007). All the isochrones are corrected for the 
adopted distance and reddening. The probable cluster members (shown with red circles in top panel of Fig. 4) clearly show that most of these are young in the CMD. As can be seen, there is a clear color gap in the cluster CMD (along the 10 Myr isochrone) that seems to separate a young stellar population 
(toward the right side in the CMD) from the likely field stars toward the cluster direction. A similar 
separation of PMS population on the CMD is also observed by Panwar et al. (2018) for the young cluster 
Berkeley 59.

There may be significant contamination by the field population along the line of sight (e.g., Sung \& Bessell 2010; Sung et al. 2013) that can also contaminate our CMD. Therefore, to further examine the distribution of our proposed cluster population, we used the young stars selected based on the X-ray and IR emission properties from Caramazza et al. (2008). 
We constructed the $V$ / ($V$ - $I$) CMD for the {\it Spitzer}-IRAC and Chandra X-ray identified stars (see Fig. 4, bottom panel). In Fig. 4 (bottom panel), the red filled circle shows the location of the Class~{\sc i} object, magenta squares represent the Class {\sc ii} objects, and blue triangles 
represent the Class {\sc iii} objects from Caramazza et al.(2008). 

By examining the locations of these YSOs in the CMD, we found that most of the YSOs are 
located to the right of the 10 Myr PMS isochrone of Siess et al. (2000), 
confirming that the contamination due to field stars in the PMS population of the cluster region is insignificant. 
A majority of the 
YSOs selected based on the IR and X-ray properties by Caramazza et al. (2008) are young low-mass stars. 

To examine the evolutionary status of the ambiguous candidate members (without 5.8 and 8.0 $\mu$m photometry) of the 
Caramazza et al. (2008) and X-ray sources from Caramazza et al. (2012), we placed these sources on the optical CMD. 
In Fig. 4 (bottom panel), cross symbols represent the optical counterparts of the X-ray sources detected by the $Chandra$ 
observations (Caramazza et al.2012) that were not cataloged by Caramazza et al. (2008) as YSOs. In contrast, the star symbols represent the ambiguous candidate members of NGC~1893 cataloged 
in Table 4 by Caramazza et al. (2008). The positions of these sources on the optical CMD suggest that a majority of them are young 
low-mass stars. The positions of these young stars alongwith the $V$ and $I$ band magnitudes are listed in Table 1.  

Comparing the distribution of all stars with the {\it Spitzer} and X-ray identified 
YSOs in the $V$/($V$ - $I$) CMD, we suggest that the stars located toward the right of the 10 Myr isochrone may belong to the PMS population of the cluster NGC~1893.
However, deep optical observations of a nearby control field or spectroscopic observations of the PMS population are necessary for the confirmation.
\begin{figure}
\centering
\includegraphics[scale = 0.64, trim = 0 0 0 0, clip]{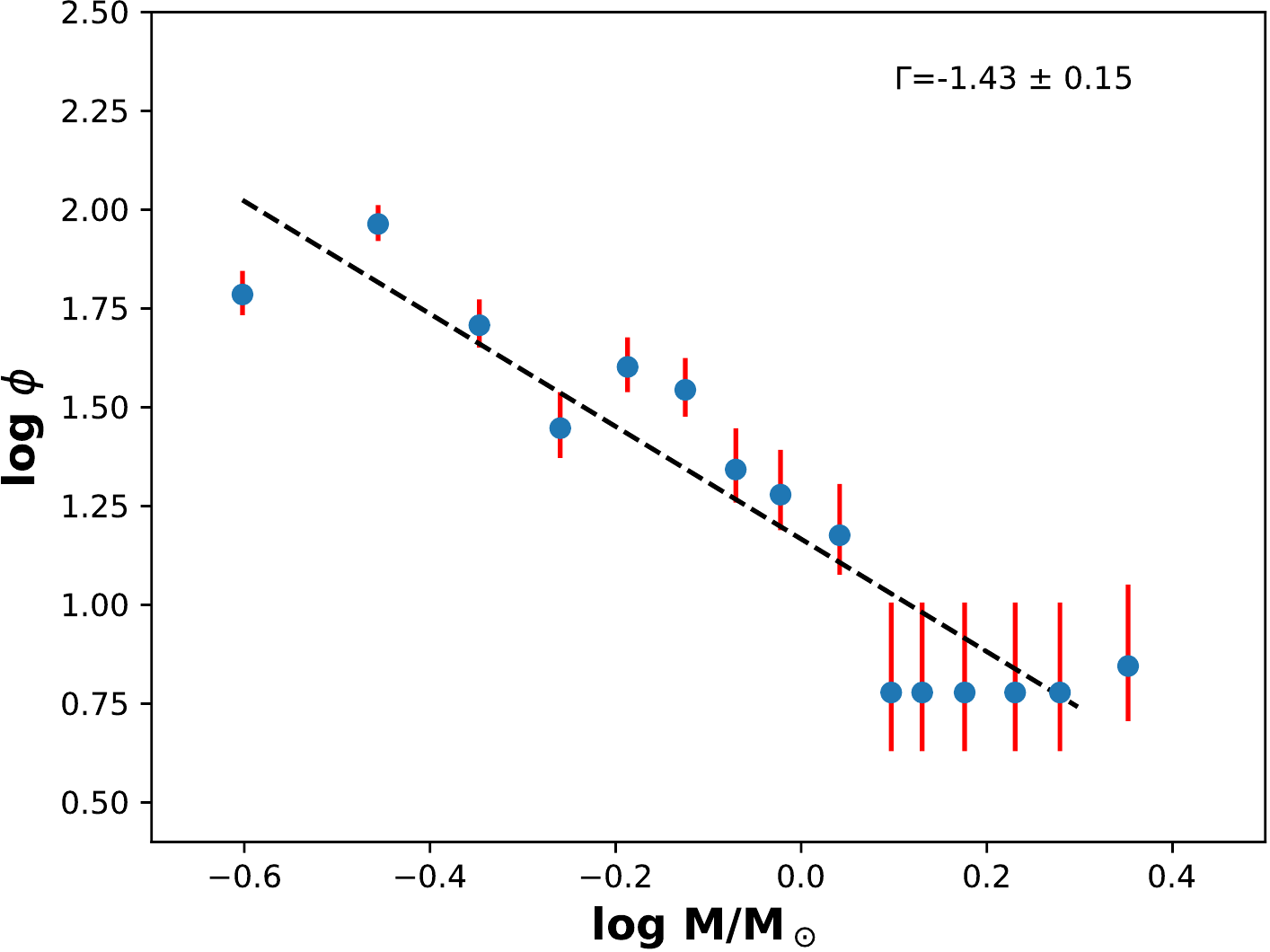}
	\caption{The MF of the YSOs in the mass range (0.2$\le$M/M$_\odot$$\le$2.5), derived from the optical data. The error bars represent $\pm$ $\sqrt{N}$ errors. The dashed line shows the least-squares fit to the mass ranges described in the text. The value of the slope obtained is given in the figure.}
        \label{fig5}
\end{figure}
\begin{figure*}
\centering
\includegraphics[scale = 0.56, trim = 0 0 0 0, clip]{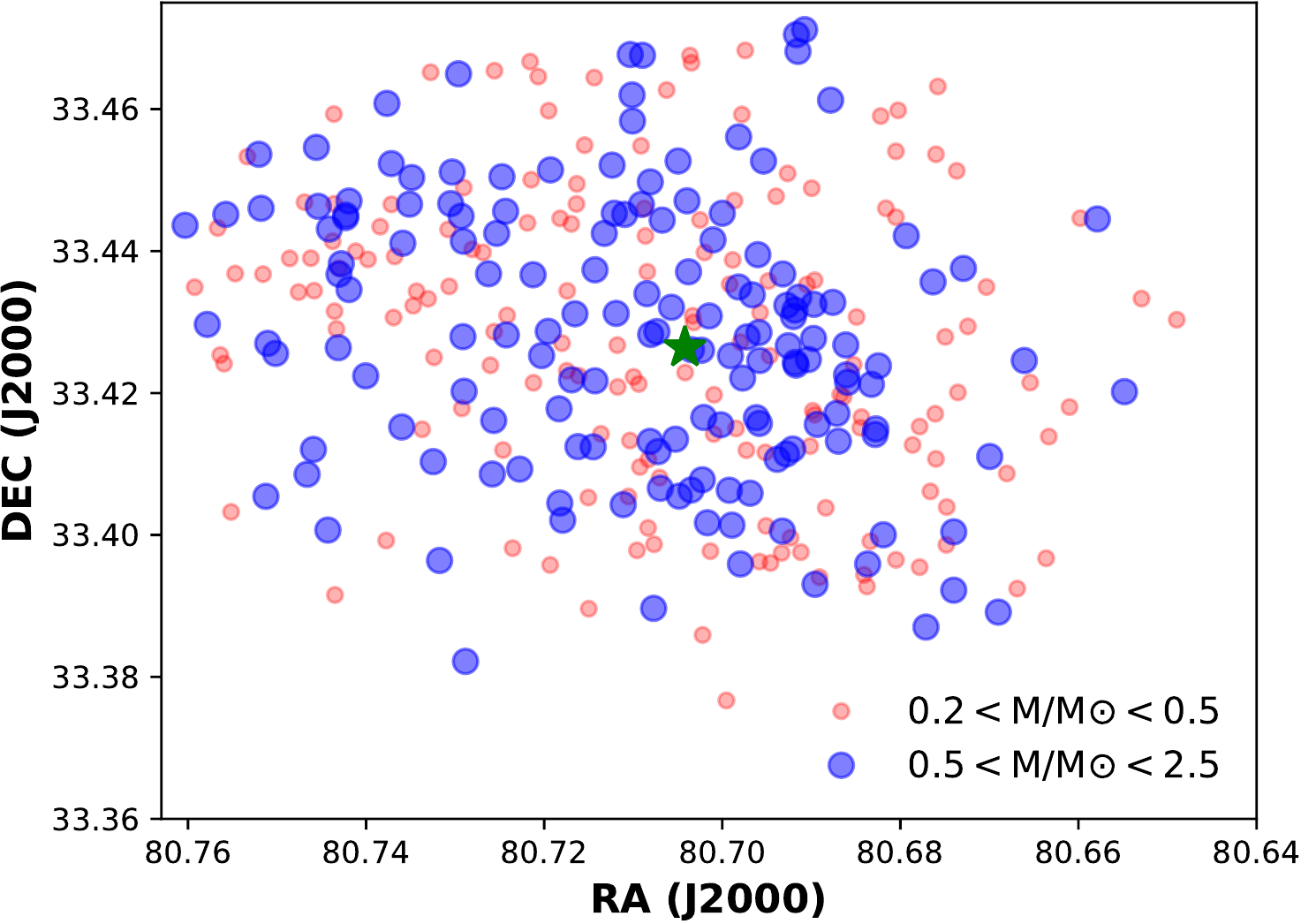}
\includegraphics[scale = 0.56, trim = 0 0 0 0, clip]{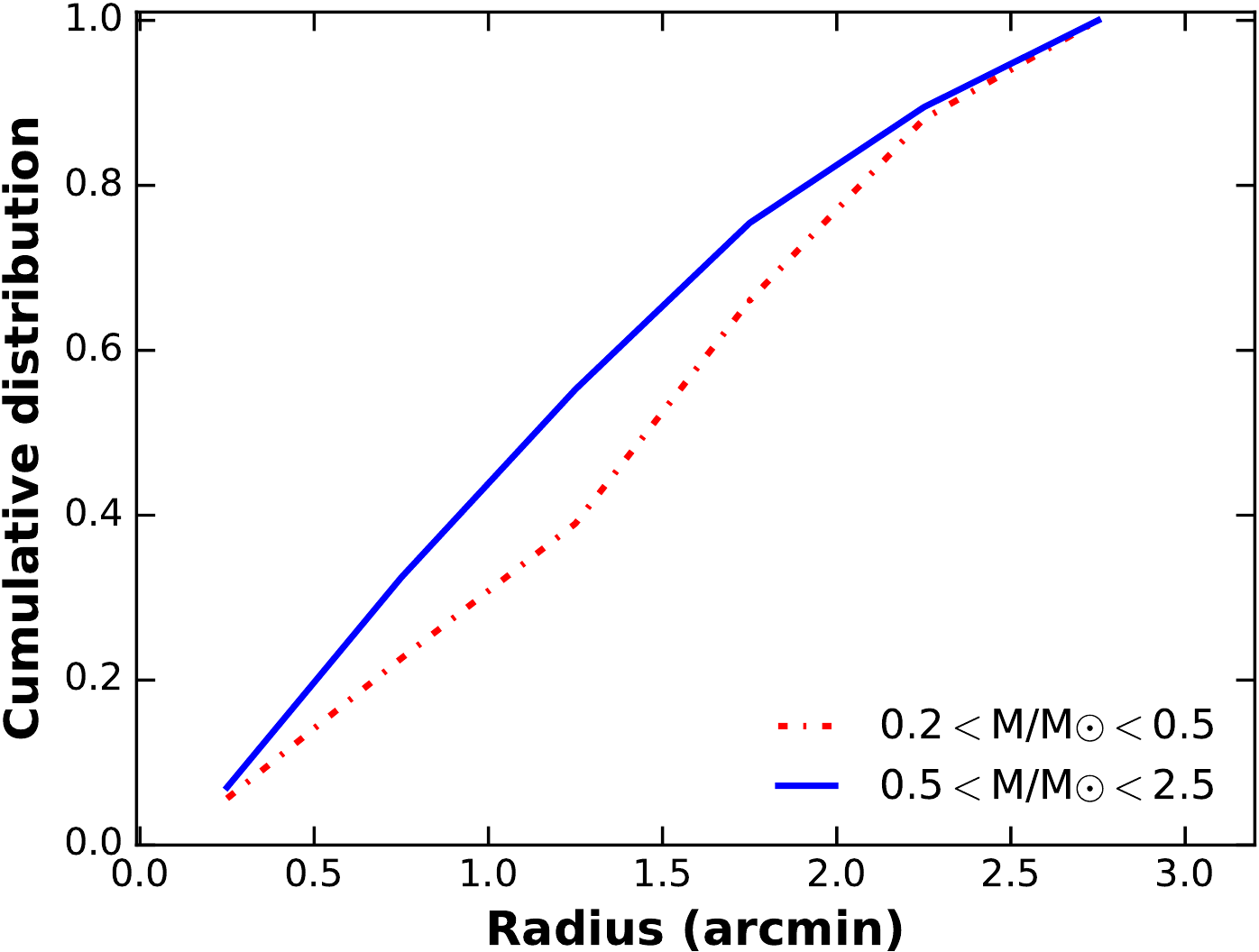}
	\caption{Left panel: Spatial distribution of the YSO candidates within the 3 arcmin radius around the cluster center (shown with green star symbol). Right panel: Cumulative radial distribution of the young stars 
	in two mass bins.}
        \label{fig6}
\end{figure*}
\subsection{Mass Function of the Young Stars}
The distribution of the stellar masses in a star cluster at a star formation event is termed as the IMF of that star cluster. 
Young star clusters (age $<$ 10 Myr) are the particularly suited for the IMF studies. They are too young to loose a significant number of members 
either due to dynamical evolution or stellar evolution. Hence, their MFs can be considered IMFs. The variation of the MF 
is an important tool that gives clues 
to the physical conditions of star formation processes (e.g., Bate 2009). The MF is defined as the number of stars per unit logarithmic mass interval, 
and is generally represented by a power law with a slope,

$\Gamma$ = d log N(log m)/d log m,

where N(log m) is the distribution of the number of stars per unit logarithmic mass interval. Observational results for the most of the star clusters in the solar neighbourhood suggest MF slopes similar to that 
given by Salpeter (1955), that is, $\Gamma$ = -1.35.

Here, we used the optical CMD to count the number of stars in different mass bins, which is shown in the lower panel of Fig. 4 along with isochrones and evolutionary tracks. 
Since our YSO sample is complete down to 0.2 M$_\odot$, for the MF study we have taken only those YSOs that have masses in the range of  
0.2$\le$ M/M$_\odot$$\le$2.5. The mass distribution of our YSO sample has a best-fitting slope, $\Gamma$ = -1.43 $\pm$ 0.15 (see Fig. 5), similar to the Salpeter value. 
Our YSO MF appears consistent with those reported for other active star-forming regions, for example, the MF of a YSO sample 
(with masses $>$0.2 M$_\odot$) derived by Erickson et al. (2011), the YSO MF slopes of the star-forming regions W51 A 
($\Gamma$ = -1.17 $\pm$ 0.26) and W51 B ($\Gamma$ = -1.32 $\pm$ 0.26) by Kang et al. (2009) 
and the YSO MF slope of the young cluster IC~1805 ($\Gamma$ = -1.23 $\pm$ 0.14) derived by Panwar et al. (2017).

The MF slope below $\sim$1 M$_\odot$ show a flattening. It has already been pointed out earlier by many groups 
(see e.g., Kroupa 2002; Chabrier 2003, 2005; Ojha et al. 2009) that for the masses above $\sim$1 M$_\odot$, the MF can generally be approximated by a declining power law with a slope similar to the Salpeter, whereas for the masses below $\sim$1 M$_\odot$ the distribution becomes flatter, 
 and turns down at the lowest stellar masses. 

\subsection{Indication for the Mass Segregation}
Although, there are extensive studies on mass segregation in star clusters, only a few focused on the mass segregation among low-mass stars (e.g., Andersen et al. 2011, Panwar et al. 2018).  Sharma et al. (2007) studied the mass segregation in the young cluster NGC 1893 in the mass range 5.5$\le$M/M$_\odot$$\le$17.7 
and found that the high-mass stars are more centrally concentrated than their lower-mass siblings. In the present work, we investigated the mass segregation in low-mass stars 
by dividing our low-mass young stars, that are located within the 3 arcmin radius around the cluster center, into two mass groups, 0.2$\le$M/M$_\odot$$\le$0.5 and 
 0.5$\le$M/M$_\odot$$\le$2.5. Fig. 6 (left panel) shows the spatial distribution of the stars from these two different groups. Toward the center of the cluster, most of the 
 YSOs seem to be relatively massive. The cumulative distribution of the young stars as a function of radial distance from the cluster center in two different mass groups (see Fig. 6, right panel) 
 also suggest that more massive stars (0.5$\le$M/M$_\odot$$\le$2.5) tend to lie toward the cluster center, which indicates the mass segregation in the central portion of the cluster region. As the estimated relaxation time for the cluster is very large compared to the age of the cluster, the 
 observed mass segregation in the cluster may be primordial in nature.
\begin{table*}
\caption{Positions and photometric magnitudes of new YSO candidates.} 
\small
\begin{tabular}{ccccc}
\hline
	$Id$ &$\alpha_{(2000)}$ & $\delta_{(2000)}$ & V $\pm$ eV& I $\pm$ eI\\
\hline
  1 & 80.65629 & 33.36719 & 24.45 $\pm$ 0.18 & 20.76 $\pm$ 0.04\\
  2 & 80.73150 & 33.37947 & 23.53 $\pm$ 0.08 & 20.52 $\pm$ 0.04\\
  3 & 80.65462 & 33.38117 & 24.93 $\pm$ 0.31 & 21.38 $\pm$ 0.11\\
  4 & 80.65742 & 33.37886 & 23.75 $\pm$ 0.10 & 20.46 $\pm$ 0.03\\
  5 & 80.73238 & 33.42503 & 23.08 $\pm$ 0.06 & 19.87 $\pm$ 0.02\\
  6 & 80.64283 & 33.38447 & 22.43 $\pm$ 0.03 & 19.07 $\pm$ 0.02\\
  7 & 80.75638 & 33.42536 & 22.43 $\pm$ 0.03 & 19.37 $\pm$ 0.02\\
  8 & 80.69583 & 33.39625 & 21.77 $\pm$ 0.02 & 18.95 $\pm$ 0.01\\
  9 & 80.65854 & 33.40050 & 23.63 $\pm$ 0.09 & 20.19 $\pm$ 0.03\\
 10 & 80.74296 & 33.37753 & 22.15 $\pm$ 0.03 & 19.74 $\pm$ 0.02\\
 11 & 80.69575 & 33.43133 & 22.75 $\pm$ 0.05 & 19.63 $\pm$ 0.02\\
 12 & 80.70321 & 33.42994 & 22.88 $\pm$ 0.04 & 19.74 $\pm$ 0.02\\
 13 & 80.64454 & 33.43739 & 21.73 $\pm$ 0.02 & 19.20 $\pm$ 0.02\\
 14 & 80.71604 & 33.44136 & 24.32 $\pm$ 0.15 & 20.67 $\pm$ 0.03\\
 15 & 80.72604 & 33.42392 & 23.23 $\pm$ 0.06 & 20.16 $\pm$ 0.02\\
 16 & 80.74096 & 33.45031 & 23.83 $\pm$ 0.10 & 20.55 $\pm$ 0.04\\
 17 & 80.69533 & 33.36542 & 22.09 $\pm$ 0.02 & 19.37 $\pm$ 0.02\\
 18 & 80.70000 & 33.44533 & 20.81 $\pm$ 0.01 & 18.57 $\pm$ 0.01\\
 19 & 80.69396 & 33.44772 & 22.76 $\pm$ 0.04 & 19.55 $\pm$ 0.02\\
 20 & 80.68996 & 33.44886 & 21.20 $\pm$ 0.02 & 18.72 $\pm$ 0.01\\
 21 & 80.73087 & 33.44939 & 24.11 $\pm$ 0.13 & 20.48 $\pm$ 0.03\\
 22 & 80.72146 & 33.45006 & 22.24 $\pm$ 0.03 & 19.38 $\pm$ 0.02\\
 23 & 80.65863 & 33.45064 & 23.40 $\pm$ 0.08 & 20.58 $\pm$ 0.04\\
 24 & 80.75554 & 33.46075 & 22.32 $\pm$ 0.03 & 19.85 $\pm$ 0.02\\
 25 & 80.67600 & 33.45364 & 23.09 $\pm$ 0.06 & 19.93 $\pm$ 0.03\\
 26 & 80.72067 & 33.46458 & 23.43 $\pm$ 0.07 & 20.06 $\pm$ 0.03\\
 27 & 80.68438 & 33.41661 & 20.44 $\pm$ 0.01 & 18.11 $\pm$ 0.01\\
 28 & 80.69267 & 33.45094 & 23.02 $\pm$ 0.06 & 19.63 $\pm$ 0.02\\
 29 & 80.72354 & 33.42625 & 23.71 $\pm$ 0.10 & 20.21 $\pm$ 0.03\\
 30 & 80.71304 & 33.46053 & 23.90 $\pm$ 0.14 & 20.32 $\pm$ 0.03\\
 31 & 80.63804 & 33.46564 & 21.61 $\pm$ 0.03 & 19.20 $\pm$ 0.02\\
 32 & 80.69742 & 33.46828 & 23.49 $\pm$ 0.11 & 20.16 $\pm$ 0.03\\
 33 & 80.68633 & 33.38081 & 24.43 $\pm$ 0.18 & 21.27 $\pm$ 0.07\\
 34 & 80.72071 & 33.38917 & 23.47 $\pm$ 0.08 & 20.05 $\pm$ 0.03\\
 35 & 80.68033 & 33.40683 & 24.32 $\pm$ 0.19 & 20.39 $\pm$ 0.03\\
 36 & 80.71533 & 33.40808 & 24.06 $\pm$ 0.13 & 20.41 $\pm$ 0.03\\
 37 & 80.73371 & 33.41489 & 22.19 $\pm$ 0.02 & 19.08 $\pm$ 0.01\\
 38 & 80.70925 & 33.40958 & 23.03 $\pm$ 0.06 & 19.89 $\pm$ 0.02\\
 39 & 80.71037 & 33.41331 & 23.66 $\pm$ 0.10 & 20.55 $\pm$ 0.03\\
 40 & 80.67242 & 33.42939 & 21.02 $\pm$ 0.02 & 18.50 $\pm$ 0.01\\
 41 & 80.69050 & 33.43528 & 23.31 $\pm$ 0.07 & 20.49 $\pm$ 0.04\\
 42 & 80.68492 & 33.43072 & 23.11 $\pm$ 0.07 & 19.88 $\pm$ 0.02\\
 43 & 80.73079 & 33.44306 & 23.13 $\pm$ 0.06 & 19.94 $\pm$ 0.02\\
 44 & 80.70250 & 33.44439 & 21.86 $\pm$ 0.02 & 19.06 $\pm$ 0.01\\
 45 & 80.65404 & 33.37956 & 23.95 $\pm$ 0.14 & 20.82 $\pm$ 0.05\\
 46 & 80.71221 & 33.36967 & 19.96 $\pm$ 0.01 & 17.82 $\pm$ 0.01\\
 47 & 80.70875 & 33.44600 & 21.99 $\pm$ 0.06 & 19.00 $\pm$ 0.02\\
 48 & 80.70008 & 33.36347 & 18.88 $\pm$ 0.01 & 16.98 $\pm$ 0.01\\
 49 & 80.70225 & 33.40778 & 19.80 $\pm$ 0.01 & 17.59 $\pm$ 0.01\\
 50 & 80.67783 & 33.41528 & 22.42 $\pm$ 0.04 & 19.57 $\pm$ 0.02\\
 51 & 80.73717 & 33.44658 & 22.15 $\pm$ 0.03 & 19.26 $\pm$ 0.02\\
 52 & 80.68675 & 33.41978 & 22.42 $\pm$ 0.04 & 19.73 $\pm$ 0.02\\
 53 & 80.68450 & 33.41511 & 21.30 $\pm$ 0.02 & 18.81 $\pm$ 0.01\\
 54 & 80.69863 & 33.44711 & 20.77 $\pm$ 0.01 & 18.13 $\pm$ 0.01\\
 55 & 80.74375 & 33.44139 & 23.61 $\pm$ 0.08 & 20.66 $\pm$ 0.04\\
\hline 
\end{tabular}
\begin{tabular}{ccccc}
\hline 
	$Id$ &$\alpha_{(2000)}$ & $\delta_{(2000)}$ & V $\pm$ eV& I $\pm$ eI\\
\hline
 56 & 80.69458 & 33.39606 & 22.87 $\pm$ 0.05 & 19.85 $\pm$ 0.02\\
 57 & 80.71696 & 33.44383 & 23.10 $\pm$ 0.05 & 20.21 $\pm$ 0.03\\
 58 & 80.72688 & 33.43981 & 22.68 $\pm$ 0.05 & 20.06 $\pm$ 0.03\\
 59 & 80.70908 & 33.44658 & 18.97 $\pm$ 0.01 & 17.41 $\pm$ 0.01\\
 60 & 80.73842 & 33.44344 & 21.17 $\pm$ 0.02 & 18.72 $\pm$ 0.01\\
 61 & 80.68525 & 33.42400 & 23.06 $\pm$ 0.05 & 20.10 $\pm$ 0.03\\
 62 & 80.68050 & 33.44478 & 22.40 $\pm$ 0.03 & 19.52 $\pm$ 0.02\\
 63 & 80.68983 & 33.41750 & 21.68 $\pm$ 0.02 & 18.75 $\pm$ 0.01\\
 64 & 80.69846 & 33.41500 & 21.70 $\pm$ 0.02 & 18.97 $\pm$ 0.01\\
 65 & 80.70829 & 33.41067 & 20.86 $\pm$ 0.02 & 18.37 $\pm$ 0.01\\
 66 & 80.70017 & 33.41553 & 20.31 $\pm$ 0.01 & 18.14 $\pm$ 0.01\\
 67 & 80.69579 & 33.42458 & 20.04 $\pm$ 0.01 & 17.94 $\pm$ 0.01\\
 68 & 80.70346 & 33.42611 & 19.88 $\pm$ 0.01 & 17.81 $\pm$ 0.01\\
 69 & 80.71450 & 33.41242 & 20.53 $\pm$ 0.01 & 18.42 $\pm$ 0.01\\
 70 & 80.68837 & 33.40383 & 22.59 $\pm$ 0.07 & 19.41 $\pm$ 0.02\\
 71 & 80.69200 & 33.43078 & 19.43 $\pm$ 0.02 & 17.53 $\pm$ 0.01\\
 72 & 80.71504 & 33.40528 & 21.51 $\pm$ 0.02 & 19.08 $\pm$ 0.01\\
 73 & 80.70096 & 33.41422 & 22.14 $\pm$ 0.03 & 19.11 $\pm$ 0.01\\
 74 & 80.71096 & 33.44517 & 20.65 $\pm$ 0.01 & 18.41 $\pm$ 0.01\\
 75 & 80.72900 & 33.42025 & 20.06 $\pm$ 0.01 & 17.91 $\pm$ 0.01\\
 76 & 80.69771 & 33.42211 & 19.71 $\pm$ 0.01 & 17.65 $\pm$ 0.01\\
 77 & 80.68375 & 33.39275 & 22.75 $\pm$ 0.04 & 20.14 $\pm$ 0.03\\
 78 & 80.70804 & 33.42825 & 17.89 $\pm$ 0.01 & 16.35 $\pm$ 0.01\\
 79 & 80.70304 & 33.37025 & 21.33 $\pm$ 0.02 & 18.85 $\pm$ 0.01\\
 80 & 80.70808 & 33.44975 & 19.61 $\pm$ 0.02 & 17.50 $\pm$ 0.01\\
 81 & 80.75825 & 33.45269 & 23.20 $\pm$ 0.07 & 20.34 $\pm$ 0.03\\
 82 & 80.70363 & 33.46756 & 22.68 $\pm$ 0.05 & 19.91 $\pm$ 0.02\\
 83 & 80.68200 & 33.37367 & 23.22 $\pm$ 0.07 & 19.90 $\pm$ 0.02\\
 84 & 80.68908 & 33.39406 & 23.13 $\pm$ 0.06 & 20.14 $\pm$ 0.03\\
 85 & 80.69333 & 33.39747 & 21.93 $\pm$ 0.02 & 18.92 $\pm$ 0.01\\
 86 & 80.74046 & 33.45139 & 24.93 $\pm$ 0.27 & 21.40 $\pm$ 0.07\\
 87 & 80.68050 & 33.45406 & 22.89 $\pm$ 0.05 & 20.09 $\pm$ 0.03\\
 88 & 80.69779 & 33.45925 & 22.97 $\pm$ 0.06 & 20.06 $\pm$ 0.02\\
 89 & 80.74229 & 33.44472 & 18.74 $\pm$ 0.04 & 17.00 $\pm$ 0.01\\
 90 & 80.69917 & 33.43533 & 22.44 $\pm$ 0.04 & 19.44 $\pm$ 0.05\\
 91 & 80.71362 & 33.41425 & 22.92 $\pm$ 0.05 & 20.22 $\pm$ 0.03\\
 92 & 80.74192 & 33.43458 & 21.00 $\pm$ 0.02 & 18.70 $\pm$ 0.02\\
 93 & 80.69258 & 33.42667 & 20.86 $\pm$ 0.04 & 18.57 $\pm$ 0.01\\
 94 & 80.71742 & 33.43439 & 22.40 $\pm$ 0.04 & 19.73 $\pm$ 0.02\\
 95 & 80.71613 & 33.42244 & 22.28 $\pm$ 0.03 & 19.74 $\pm$ 0.02\\
 96 & 80.70333 & 33.43089 & 20.87 $\pm$ 0.01 & 18.27 $\pm$ 0.01\\
 97 & 80.73067 & 33.43503 & 21.46 $\pm$ 0.02 & 18.96 $\pm$ 0.01\\
 98 & 80.67933 & 33.44217 & 19.81 $\pm$ 0.01 & 17.87 $\pm$ 0.01\\
 99 & 80.74229 & 33.44506 & 20.02 $\pm$ 0.02 & 17.86 $\pm$ 0.01\\
100 & 80.69483 & 33.43578 & 23.22 $\pm$ 0.06 & 20.21 $\pm$ 0.03\\
101 & 80.69554 & 33.44658 & 24.25 $\pm$ 0.19 & 20.62 $\pm$ 0.04\\
102 & 80.67579 & 33.46319 & 23.33 $\pm$ 0.09 & 20.53 $\pm$ 0.04\\
103 & 80.65292 & 33.43331 & 22.72 $\pm$ 0.05 & 19.25 $\pm$ 0.02\\
104 & 80.69171 & 33.42394 & 18.96 $\pm$ 0.02 & 16.90 $\pm$ 0.01\\
105 & 80.68162 & 33.44603 & 20.76 $\pm$ 0.01 & 18.30 $\pm$ 0.01\\
106 & 80.68975 & 33.42769 & 18.88 $\pm$ 0.01 & 16.98 $\pm$ 0.01\\
107 & 80.71637 & 33.44664 & 22.21 $\pm$ 0.04 & 19.52 $\pm$ 0.02\\
108 & 80.71179 & 33.42675 & 21.53 $\pm$ 0.05 & 18.89 $\pm$ 0.01\\
109 & 80.69796 & 33.42717 & 22.94 $\pm$ 0.05 & 19.77 $\pm$ 0.02\\
110 & 80.69467 & 33.42525 & 20.74 $\pm$ 0.02 & 18.24 $\pm$ 0.01\\
\hline 
\end{tabular}
\label{tab1}
\end{table*}
\section{Summary \& conclusions}
Here, we present the results of our deep optical ($VI$) observations of the central portion of the cluster NGC~1893. Thanks to the excellent observing conditions and spatial resolution of 4K$\times$4K CCD $IMAGER$, 
{present data are $\sim$ 3 mag deeper than most of the previous studies.}
\begin{itemize}
	\item Considering a distance of $\sim$ 3.25 kpc and reddening E($B$ - $V$) 
of $\sim$ 0.4 mag for the cluster, our optical data are deep enough to reveal the stars below $\sim$ 0.2 M$_\odot$. 
	\item We found counterparts of $\sim$ 450 YSOs, candidate members and X-ray sources in our optical catalog. 
	\item We estimated the membership probability of the stars in the cluster region using the Gaia EDR3. The parallax values of the stars with high membership probability and good parallax measurements 
($\varpi$/$\sigma$$_\varpi$ $>$ 5) are used to estimate the cluster distance. The estimated distance of $\sim$ 3.30 kpc is in good agreement with the photometric distance estimate for the cluster reported by Sharma et al. (2007). 
\item The locations of the young stellar candidates in the 
$V$/($V$ - $I$) CMD show that most of them have ages $<$ 10 Myr. Comparing the CMDs for all stars and young members in the cluster region, 
we suggest an insignificant contribution of the field stars in the PMS zone of the cluster CMD and young stellar population of the cluster region can be identified using the CMD. 
		\item We also found that most of the candidate members having no classification flag in Caramazza et al. (2008) and X-ray sources in Caramazza et al. (2012) also occupy the locations 
of PMS stars in CMD. 
		\item Based on the present optical observations, we identified $\sim$ 425 young stars in the central portion of the cluster NGC~1893 and $\sim$ 110 of these were new. Our results also demonstrate that present optical $VI$ observations with the 4K$\times$4K CCD $IMAGER$ reveal faint low-mass stars ($V$ $\sim$ 25 mag at 60\% reflectivity of the primary mirror) in the cluster NGC~1893. 
		\item The MF of our YSO sample has a power-law index 
of -1.43 $\pm$ 0.15, close to the Salpeter value (-1.35) and reported for other star forming regions. The spatial distribution of the YSOs as a function of mass suggest that 
		toward the cluster center, most of the stars are relatively massive. 
\end{itemize}

\section*{Acknowledgements}
We thank the referee for valuable comments that significantly improved the manuscript. We are thankful to the DTAC and staff of the 3.6-m DOT which is operated by ARIES. 
This work has made use of data from the European Space Agency (ESA) mission {\it Gaia} (https://www.cosmos.esa.int/gaia), processed by the {\it Gaia}
Data Processing and Analysis Consortium. Funding for the DPAC
has been provided by national institutions, in particular the institutions
participating in the {\it Gaia} Multilateral Agreement.

\vspace{-1em}

%
%
%

\end{document}